\documentclass[%
 reprint,
superscriptaddress,
amsmath,amssymb,
aip,
]{revtex4-1}

\usepackage{graphicx}
\usepackage{dcolumn}
\usepackage{bm}
\usepackage{hyperref}

\usepackage{xcolor}
\begin{document}

\title{Acceleration sensing with magnetically levitated oscillators above a superconductor}

\author{Chris Timberlake}
\email{ct10g12@soton.ac.uk}
\affiliation{Department of Physics and Astronomy, University of Southampton, Southampton SO17 1BJ, United Kingdom
}

\author{Giulio Gasbarri}
\affiliation{Department of Physics and Astronomy, University of Southampton, Southampton SO17 1BJ, United Kingdom
}

\author{Andrea Vinante}
\affiliation{Department of Physics and Astronomy, University of Southampton, Southampton SO17 1BJ, United Kingdom
}

\author{Ashley Setter}
\affiliation{Department of Physics and Astronomy, University of Southampton, Southampton SO17 1BJ, United Kingdom
}

\author{Hendrik Ulbricht}
\email{h.ulbricht@soton.ac.uk}
\affiliation{Department of Physics and Astronomy, University of Southampton, Southampton SO17 1BJ, United Kingdom
}

\date{\today}

\begin{abstract}

We experimentally demonstrate stable trapping of a permanent magnet sphere above a lead superconductor, in vacuum pressures of $4 \times 10^{-8}$~mbar. The levitating magnet behaves as a harmonic oscillator, with frequencies in the 4-31~Hz range detected, and shows promise to be an ultrasensitive acceleration sensor. We directly apply an acceleration to the magnet with a current carrying wire, which we use to measure a background noise of $\sim 10^{-10} \ \text{m}/\sqrt{\text{Hz}}$ at 30.75~Hz frequency. With current experimental parameters, we find an acceleration sensitivity of $S_a^{1/2} = 1.2 \pm 0.2 \times 10^{-10} \ \text{g}/\sqrt{\text{Hz}}$, for a thermal noise limited system. By considering a 300~mK environment, at a background helium pressure of $1 \times 10^{-10}$~mbar, acceleration sensitivities of $S_a^{1/2} \sim 3 \times 10^{-15} \ \text{g}/\sqrt{\text{Hz}}$ could be possible with ideal conditions and vibration isolation. To feasibly measure with such a sensitivity, feedback cooling must be implemented.

\end{abstract}

\maketitle

The ability to detect extremely small forces and accelerations has a diverse range of applications within science and technology, including uses in magnetic resonance force microscopy~\cite{Rugar2004, Degen2009, Moser2013}, detection of gravitational waves~\cite{Abbott2016}, measuring short range Casimir forces~\cite{Klimchitskaya2009}, gravimetry~\cite{Marson2012} and measuring gravitational fields of small source masses~\cite{Schmole2016}. Such systems could also be utilized to test fundamental physics, such as testing collapse models which predict extensions to standard quantum mechanics~\cite{Bassi2003, Bassi2013}, as well as searching for non-Newtonian corrections to our understanding of gravity~\cite{Geraci2015}. State-of-the-art force sensors, based on clamped mechanical resonators, have reached force sensitivities of $\sim 10^{-21} \ \text{N}/\sqrt{\text{Hz}}$~\cite{Moser2014} in cryogenic environments and $\sim 10^{-17} \ \text{N}/\sqrt{\text{Hz}}$~\cite{Norte2016, Reinhardt2016} at room temperature. These mechanical resonators are limited in their sensitivity due to the dissipation to the clamping losses. A natural solution to avoid such losses is to levitate the resonator. Indeed, optically levitated dielectric particles~\cite{Vovrosh2017, Gieseler2012, Barker2010, Romero-Isart2010, Li2010, Kiesel2013, Chang2010} have shown high quality factors, with force sensitivities of $\sim 10^{-20} \ \text{N}/\sqrt{\text{Hz}}$ achieved~\cite{Hempston2017, Gieseler2013} and short range interactions between dielectric surfaces and the particle investigated~\cite{Winstone2018, Diehl2018}. For acceleration measurements, the best performances are obtained with massive systems; impressive sensitivities of $< 10^{-15} \ \text{g}/\sqrt{\text{Hz}}$ in the LISA Pathfinder in-flight experiment~\cite{Armano2016} have been demonstrated. For commercial uses, superconducting gravimeters, which levitate a centimeter sized type-II superconductor, have achieved acceleration sensitivities of $\sim 10^{-10} \ \text{g}/\sqrt{\text{Hz}}$~\cite{Goodkind1999}. 

In principle, magnetically levitated oscillators could provide the most environmentally isolated oscillators; the trapping mechanism is passive, whereas other levitation systems require active fields which limit the $Q$-factor and the temperature attainable~\cite{Vinante2019}. Due to this promise, magnetically levitated oscillators have been proposed as a route to observing macroscopic superposition states~\cite{Romero-Isart2012, Cirio2012, Pino2018, Johnsson2016}, as well as for force and inertial sensing~\cite{Prat-Camps2017}, magnetometry~\cite{Kimball2016} and gravimetry~\cite{Johnsson2016}. Experimentally, diamagnetic microparticles have been levitated with permanent magnets, and had its harmonic centre of mass motion cooled to sub-K~\cite{Hsu2016} and sub-mK~\cite{Slezak2018, klahold2019} temperatures in vacuum conditions. Additionally, ferromagnetic microparticles have been levitated above type-II superconductors~\cite{Wang2019}, which show promise to be ultra-sensitive torque sensors and magnetometers.    

In this paper, we experimentally demonstrate magnetic trapping in high vacuum conditions, using a single permanent magnet levitated above a type-I superconductor to generate the trapping potential. The relatively high mass, with respect to nano and micro-particles widely used in levitated optomechanics, and low frequency of our oscillator makes it a strong candidate for compact acceleration sensing, even at moderate $Q$-factors. Indeed, the evaluated acceleration sensitivity for our system is $S_a^{1/2} = 1.2 \pm 0.2 \times 10^{-10} \ \text{g}/\sqrt{\text{Hz}}$, assuming it is thermal noise limited (g is the gravitational acceleration on Earth). Further sensitivity improvements are also predicted by transferring the experimental setup to a fully cryogenic environment, reducing the temperature of the thermal bath, and removing nearby resistive metals, where energy is currently lost via eddy current dissipation. We also apply a series of known accelerations to our oscillator, and use this to estimate the background noise to be $\sim 10^{-10} \ \text{m}/\sqrt{\text{Hz}}$ at 30.75~Hz frequency. The minimum RMS acceleration we apply is $a_{\text{RMS}} = 8.2 \pm 3.3 \times 10^{-7}$~g, which is easily distinguishable from zero applied acceleration. 

\begin{figure*}[t!]
    \centering
    \includegraphics[width=0.96\textwidth]{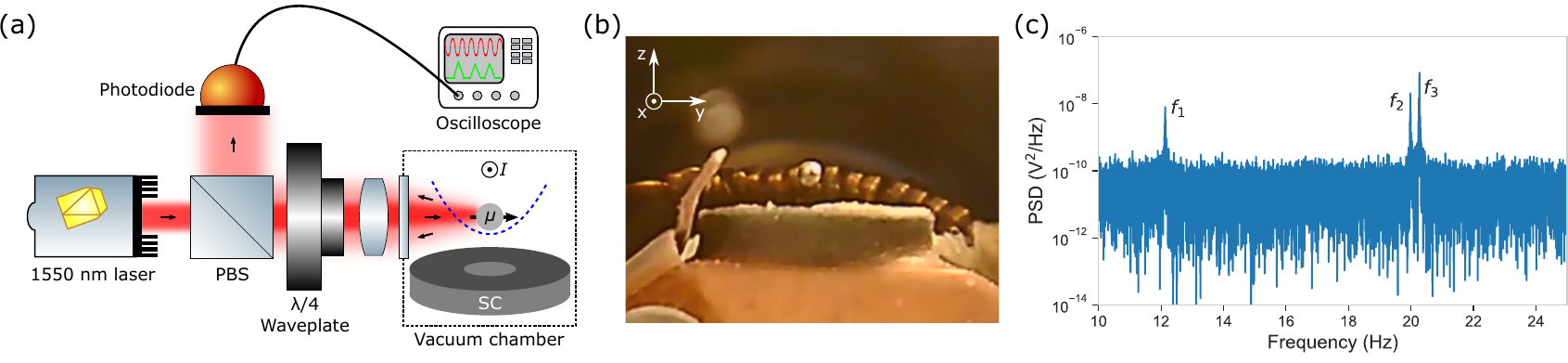}
    \caption{(a) Experimental setup. A 0.5~mm radius neodymium magnet sphere is levitated above a lead superconducting disk, in a vacuum chamber. A current carrying wire is placed close to the trap centre, allowing us to modulate the magnetic field. The lead is cooled to $\sim$ 5~K with a liquid helium continuous flow cryostat. For position detection, a 1550~nm laser is focused onto the levitating magnet, and the reflected beam is collected with a photodiode. (b) A photograph of the magnet levitating above the lead disk. (c) A typical Power Spectral Density (PSD) of the optical power reflected from the magnet. Spectrum produced from 1000 s of data, at a vacuum pressure of $4 \times 10^{-8}$~mbar.}
    \label{fig:setup}
\end{figure*}

The experimental setup consists of a room temperature, 0.5~mm radius, neodymium magnet sphere (mass $m$ = 4~mg, magnetization $M = 1.1 \times 10^{-7}$~A/m) that levitates above a superconducting lead disk with outer diameter of 10.0~mm, inner diameter of 3.0~mm, and a height of 2.5~mm, which is fixed to the cold finger of a liquid helium continuous flow cryostat used to cool the lead to $\sim$ 5~K. Lead was chosen due to having the highest critical temperature ($T_{c}\sim7.2$~K) of the type-I superconductors. Additionally, a 5.5~cm straight copper wire placed 1.3~cm above from the disk is used to modulate the magnetic field. For detection,  a 1550~nm laser (power $\sim100 \ \mu \text{W}$) is focused  onto the magnet surface, reflected, recollimated, and collected with  a photodiode. This  power  was  chosen  as  it  was found experimentally that higher powers would heat and demagnetize the magnet - resulting in the magnet being ejected from the trap. A schematic of the trapping and detection setup can be seen in Fig.~\ref{fig:setup} (a), and an image of levitation in Fig.~\ref{fig:setup} (b). The whole system is inserted into a vacuum chamber and the magnet is loaded into position with a translation stage, which is withdrawn from the levitation region prior to any measurements. In this configuration the superconducting disk is used to generate a force $\bm{F}_\text{disk} = -\frac{1}{2} (\bm{\mu \cdot  }\nabla )\bm{B_{\text{disk}}}$, where $\bm{\mu}$ is the magnetic dipole moment and $\bm{B_{\text{disk}}}$ the magnetic field induced by the superconductor, that compensates for the gravitational force $\bm{F}_{g}= -mg\bm{z}$ on the vertical axis and generates a trapping potential along the horizontal plane~\cite{Prat-Camps2017}.

The trapping potential generated by this configuration has been studied using the finite element analysis software FEniCS~\cite{AlnaesBlechta2015a}, under the assumptions of the superconducting disk is cooled in the absence of any external field, in the limit in which the London penetration depth is negligible and the system can be considered to be in a quasi-static configuration.
The simulation shows an equilibrium position and orientation of $ x_0=0.1 $~mm, $y_0=0$~mm, $z_0=3.0$~mm,  $\alpha_{0} = 0.5 \pi $ and $\beta_{0}=0.0 $ in the presence of the Earth's magnetic field\footnote{The Earth's magnetic field in Southampton, UK (at sea level) has a vertical component of 44469.57~nT and a horizontal component of 19813.59~nT}, and that the 
trapping potential, around this equilibrium, can be approximated by the harmonic potential
\begin{align}
V(\bm{r},\bm{\theta})&=
\frac{m}{2}\bm{r}\, \hat{\omega}^2\, \bm{r}^{\top}+ \frac{I}{2}\bm{\theta}\,\hat{\Omega}^2\,\bm{\theta}^{\top} +\sqrt{m I}\, \bm{\theta}\,  \hat{\kappa}\, \bm{r}^{\top},
\end{align}
with
\begin{align}
\frac{\hat{\omega}}{2\pi}&=
\left(
\begin{array}{ccc} 
 4.3& 0.0 & 0.1 \nonumber\\
 0.0&	 4.4 & 0.1 \nonumber\\
 0.1& 0.1 & 18.1 
   \end{array}
  \right) \text{Hz},\hspace{0.15cm} 
  \frac{\hat{\Omega}}{2\pi}= \left(
\begin{array}{ccc} 
 67.8& 0.1  \nonumber\\
 0.1&17.5 
   \end{array}
\right)\text{Hz},\nonumber\\
\frac{\hat{\kappa}}{(2\pi)^2}&= \left( 
\begin{array}{ccc}
65.6&0.0&158.8\nonumber\\
1.4&7.3&0.4\nonumber\\
\end{array}
\right)\text{Hz}^2,
\end{align}

where $\bm{r}=(x,y,z)$  with $x$ being the direction of the Earth's magnetic field in the horizontal plane, $y$ the transverse direction, $z$ the vertical direction, $\bm{\theta}=(\alpha,\beta)$ with $\alpha$ and $\beta$ the angles between $\bm{\mu}$ and $z$, and $\bm{\mu}$ and $x$ respectively, $m$ is the mass of the magnet and $I$ its moment of inertia. 
The associated equations of motion, where the stochastic forces, torques and magnetic losses are also considered, reads:
\begin{equation}
\left\{
\begin{split}
m(\ddot{\bm{r}}+ \hat{\omega}^{2} \bm{r}^{\top}+\hat{\gamma}_{r}\cdot\bm{\dot{r}}^{\top} +\sqrt{I m}\bm{\theta}\,\hat{\kappa}) = \bm{f},\\
I(\ddot{\bm{\theta}}+\hat{\Omega}^{2}\bm{\theta}+\hat{\gamma}_{\theta}\cdot\bm{\dot{\theta}}^{\top}+\sqrt{I m}\,\hat{\kappa}\,\bm{r}^{\top}) = \bm{\tau},\\
\end{split}\right.
\end{equation}
 where $\bm{f}$ and $\bm{\tau}$ represent the stochastic force and torque and $\gamma$ the    damping rates due to gas collisions and magnetic losses \cite{Prat-Camps2017}.
 From these equations we see that the eigenmodes of the system mixes translational and librational degrees of freedom and the associated eigenfrequencies read as $\frac{\omega_{1}}{2\pi}=67.9$~Hz, $\frac{\omega_{2}}{2\pi}=17.9$~Hz, $\frac{\omega_{3}}{2\pi}=17.5$~Hz,$\frac{\omega_{4}}{2\pi}=4.4$~Hz and  $\frac{\omega_{5}}{2\pi}=4.2$~Hz.
 
\begin{figure*}[t!]
    \includegraphics[width=0.96\textwidth]{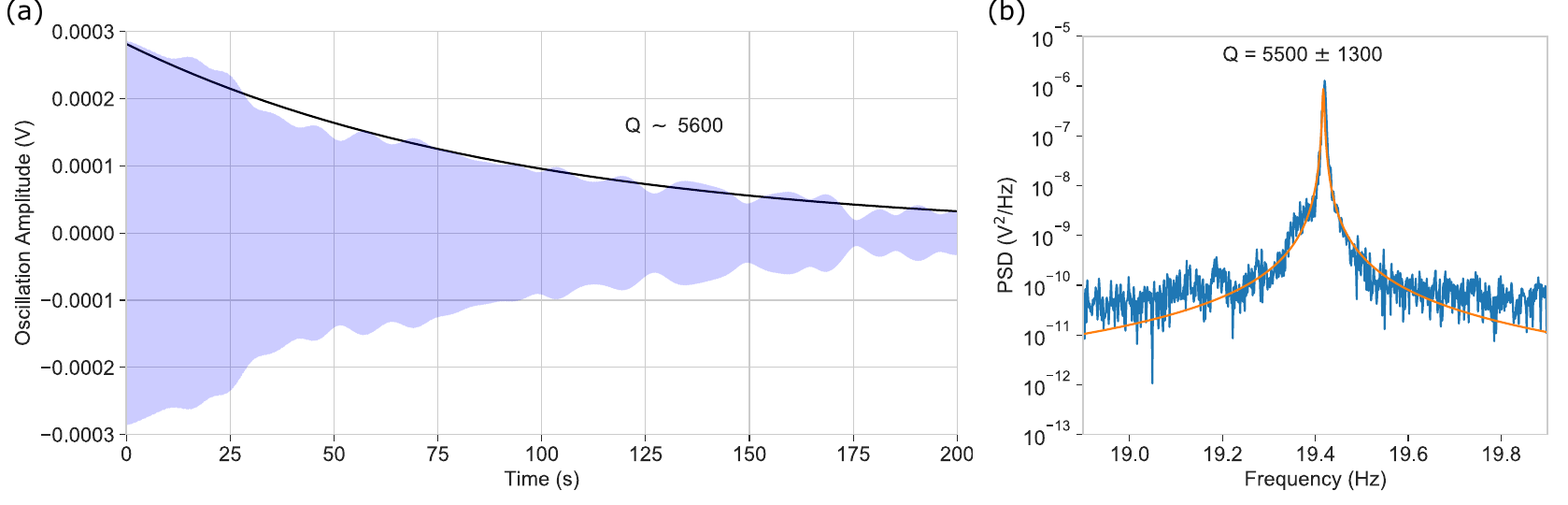}
  \caption{(a) A ring down measurement of an oscillation mode. The levitating 0.5~mm radius magnet sphere is excited by ``kicking" the vacuum chamber gently and observing the amplitude decay of the envelope. The frequency mode is isolated with the use of a bandpass filter, and the decay is fitted with an exponential function with a time constant $\tau = \text{93~s}$. (b) The $Q$-factor is extracted from fitting a Lorentzian distribution the PSD of the frequency mode. The PSD is generated with 2000~s of data, at a vacuum pressure of $8 \times 10^{-7}$~mbar.}
  \label{fig:damping}
\end{figure*} 
 
A typical frequency spectrum of the levitating magnet can be seen in Fig.~\ref{fig:setup} (c). Modes are identified as having translational components due to them being easily excitable, for instance by giving the vacuum chamber a mechanical``kick". The frequencies obtained experimentally agree with the simulation results to within an order of magnitude. Deviations are explained by variations in the experiment, including magnetization of the oscillator, uncertainties on the geometry of the lead disk, as well as nearby ferromagnetic material - such as mounting screws, temperature sensors and the vacuum chamber itself. Indeed, we see variations of trap frequencies (withing the range 4-31~Hz) from different experimental runs. Additionally, the simulation considers the Earth's magnetic field to be reflected by the superconductor, whereas in reality there may be some magnetic field frozen within the hole of the superconducting disk.

The acceleration sensitivity of a thermal noise limited system is determined by the thermal force noise on the oscillator, which is given by

\begin{equation}
    S_F = 4 k_B T m \omega_0/Q,
\label{eq:Thermal_noise}
\end{equation}

where $k_B$ is the Boltzmann constant, $T$ is the temperature of the thermal bath, $\omega_0$ is the resonance frequency and $Q$ is the quality factor \cite{Majorana1997}. The acceleration sensitivity is then

\begin{equation}
    S_a^{1/2} = \sqrt{\frac{4 k_B T \omega_0}{m Q}}.
    \label{eq:Acc_sensitivity}
\end{equation}

The $Q$-factor of the oscillator is determined by damping mechanisms in the system. Currently, the dominating damping mechanism is due to eddy current dissipation in nearby resistive metals, including the coating of the magnet itself (thickness of 12-25 $\mu\text{m}$ of nickel-copper-nickel), which is unavoidable in the current setup; other damping sources are many orders of magnitude smaller in the experimental conditions. By measuring the exponential decay time of the amplitude, after the oscillator is excited by ``kicking" the vacuum chamber, we obtain a time constant of $\tau = \text{93~s}$, which corresponds to $Q \sim 5600$. This $Q$ is in good agreement with the value $Q = 5500 \pm 1300$ extracted from the width of the Lorentzian peak in the frequency spectrum. The $Q$-factor extraction methods can be seen in Fig.~\ref{fig:damping}.   

Assuming the noise is thermally limited, the evaluated acceleration sensitivity in our experiment is $S_a^{1/2} = 1.2 \pm 0.2 \times 10^{-10} \ \text{g}/\sqrt{\text{Hz}}$; the thermal noise driving the system is due to current fluctuations in the nearby 5~K metals. This number is comparable to other acceleration sensitivities mentioned earlier, and, in principle, could be considerably lower if we put the experiment in a fully cryogenic environment, remove nearby resistive metals, and use an uncoated magnet. By removing the effects of eddy current dissipation, the mechanical damping should be dominated by gas collisions. In a low pressure regime, the mechanical damping is proportional to the gas pressure, and is given by

\begin{equation}
    \Gamma_{\text{gas}} \approx \frac{15.8 r^2 P}{m \bar{v}_{\text{gas}}},
    \label{eq:gas_damping}
\end{equation}

where $r$ is the radius of the levitated sphere, $P$ is the gas pressure and $\bar{v}_{\text{gas}}$ is the thermal velocity of the gas molecules~\cite{Beresnev1990}. 

By considering the environment generated by a 300~mK cryostat, with a helium gas pressure of $1 \times 10^{-10}$~mbar, the predicted $Q$-factor would be $Q \sim 5 \times 10^{11}$, for a thermal noise limited system. In order to achieve such a high $Q$, one has to take extreme care to avoid eddy current dissipation, and other potential dissipation sources, such as magnetic hysteresis losses, must be investigated. For such a $Q$ the acceleration sensitivity is $S_a^{1/2} \sim 3 \times 10^{-15} \ \text{g}/\sqrt{\text{Hz}}$. A cryostat with these specifications is being manufactured for our future experiments, which motivates our choice of environmental conditions. 

\begin{figure*}[t!]
    \centering
    \includegraphics[width=0.96\textwidth]{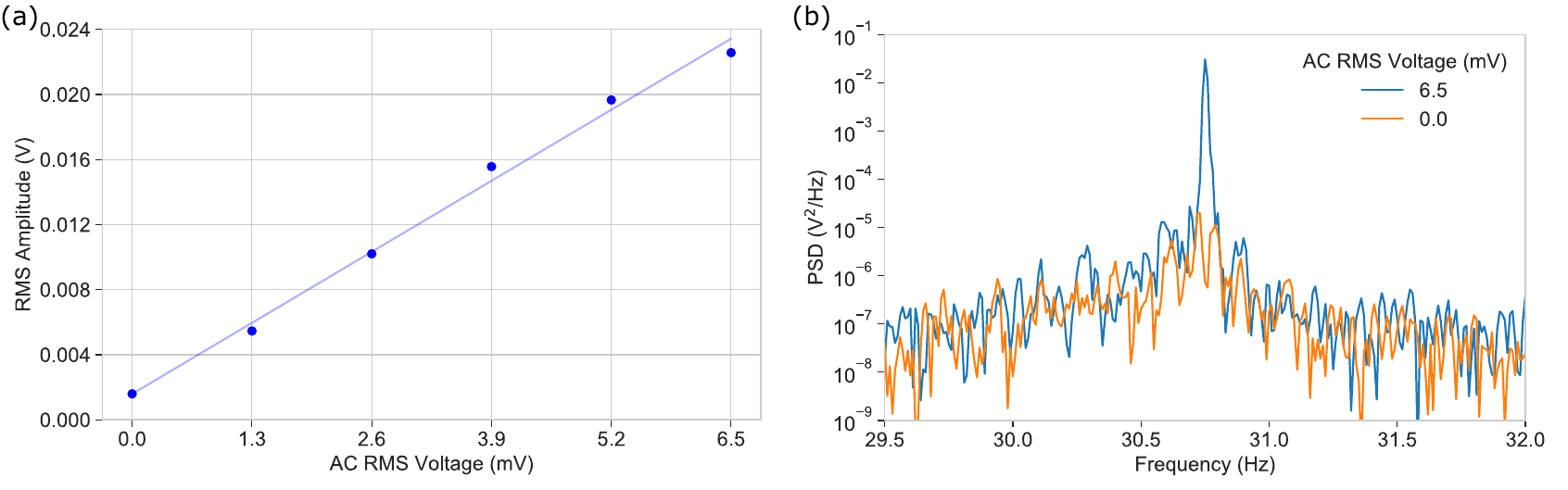}
    \caption{(a) The RMS amplitude of the oscillator as a function of applied AC RMS voltage. The amplitude increases linearly with voltage, and we can easily distinguish the response of 1.3~mV RMS voltage, compared to zero applied voltage. The frequency of the mode is $\frac{\omega_0}{2\pi} = 30.75~\text{Hz}$. (b) Comparison of the PSD of the frequency mode with zero applied voltage compared to 6.5 mV RMS. Each voltage value was applied for 100~s.}
    \label{fig:acceleration}
\end{figure*}

From a practical point of view, such an acceleration sensitivity would take an unfeasibly long time to measure, with a characteristic time scale of $2Q/\omega_0 \sim 8 \times 10^9 \ \text{s} \approx 250 \ \text{years}$. In order to preserve sensitivity, but reduce the amount of time needed to resolve the motion of the oscillator, feedback cooling must be implemented. Feedback cooling applies a damping to the oscillator, which simultaneously decreases the effective temperature of the oscillator and reduces the $Q$-factor~\cite{Geraci2010}. In our system, feedback cooling could be applied by modulating the magnetic field~\cite{Hsu2016} with the use of superconducting coils, and optimal feedback protocols could be utilized~\cite{Ferialdi2019, Tebbenjohanns2019}. In order to reach a thermally limited system, the effects of external vibrations must be mitigated. In a typical laboratory, the seismic noise is given by $\sim \lvert10^{-9}/\left(f/1 \ \text{Hz}\right)^2\rvert \ \text {m}/\sqrt{\text{Hz}}$ above 1~Hz~\cite{ze2001}, which translates to $\sim 4 \times 10^{-9} \ \text{g}/\sqrt{\text{Hz}}$ acceleration noise at $\frac{\omega_0}{2 \pi} = 19.4$~Hz. Depending on lab location, this noise contribution can differ by approximately an order of magnitude, but to achieve a thermal noise limited system, vibration damping must be implemented. In order to reach the ultimate sensitivity, such experimental issues must be overcome. 

We can directly apply a frequency-dependent acceleration to the magnet by modulating the magnetic field at the trapping position by applying a current to a nearby wire. Fig.~\ref{fig:acceleration} (a) shows the amplitude response of an oscillation mode ($\frac{\omega_0}{2\pi} = 30.75~\text{Hz}$) as a function of applied AC voltage. It can be seen that we can easily distinguish between 0.0~mV and 1.3~mV RMS voltage, and the amplitude response is linear with applied voltage. For these experiments, the oscillations were detected by measuring the shadow of the sphere produced by an incident 1550~nm laser, rather than the reflected intensity as in all other results presented. This allowed higher oscillation amplitudes to be detected compared to the reflected method described in Fig.~\ref{fig:setup} (a), at the cost of a lower detection sensitivity ($\sim 10^{-9} \ \text{m}/\sqrt{\text{Hz}}$ as opposed to $\sim 10^{-11} \ \text{m}/\sqrt{\text{Hz}}$ in the reflected detection method). Each voltage is applied for 100~s.

The force experienced by a magnetic moment is a magnetic field produced by a current carrying wire is given by $\bm{F}_\text{wire} = \nabla (\bm{\mu \cdot} \bm{B_{\text{wire}}})$, where $\bm{B_{\text{wire}}}$ is the magnetic field produced by the wire. The wire is 5.5~cm long, and the magnet levitates 1~cm below the centre of this wire. By considering the wire to be an infinitely long wire, and assuming that the magnetic moment of the magnet aligns with the Earth's magnetic field in the horizontal plane, we can use the current flowing through the wire to estimate the RMS acceleration. We then use the applied acceleration to estimate the background noise on our oscillator, which we find to be $\sim 10^{-10} \ \text{m}/\sqrt{\text{Hz}}$. This noise is slightly higher than typical seismic noise, due to the use of vacuum pumps, and would be improved with vibration isolation. Despite the high background noise, we can easily distinguish an applied RMS acceleration of $a_{\text{RMS}} = 8.2 \pm 3.3 \times 10^{-7}$~g. From this, we estimate a minimum detectable RMS acceleration of $a_{\text{min}} = 2.2 \pm 0.9 \times 10^{-7}$~g, which would double the RMS amplitude. It is worth noting that we have been conservative with our acceleration estimate, as we did not consider the vertical skew of magnetic moment angle induced by the Earth's magnetic field, which would slightly lower our acceleration estimate.

In conclusion, we have experimentally demonstrated magnetic trapping of a 0.5~mm radius neodymium magnet sphere, using superconducting lead to generate a passive trapping potential, which is stable at $4 \times 10^{-8}$~mbar background pressure. We also find an acceleration sensitivity of $S_a^{1/2} = 1.2 \pm 0.2 \times 10^{-10} \ \text{g}/\sqrt{\text{Hz}}$, with current experimental parameters, for a thermal noise limited system. Currently, the $Q$-factor is limited to around $Q \sim 10^4$ by eddy current dissipation. By removing the influence of nearby resistive metals, and placing the entire system in a 300~mK cryostat, at a background helium pressure of $1 \times 10^{-10}$~mbar, we speculate acceleration sensitivities of $S_a^{1/2} \sim 3 \times 10^{-15} \ \text{g}/\sqrt{\text{Hz}}$ could be possible for a well isolated system. In order to feasibly measure with such a sensitivity, feedback cooling must be implemented to decrease the characteristic time of the system. By applying a series of known accelerations, we also calculate our current background noise to be $\sim 10^{-10} \ \text{m}/\sqrt{\text{Hz}}$ at a frequency of 30.75~Hz. The minimum RMS acceleration we apply is $a_{\text{RMS}} = 8.2 \pm 3.3 \times 10^{-7}$~g, which we can easily distinguish from zero applied acceleration. Such a sensor could be used for gravimetry, to study gravitational effects at the small distance scale, or be used to measure gravitational attraction from extremely small masses in a Cavendish-like experiment~\cite{Cavendish1798}. 

We would like to thank M. Toro\v{s}, M. Rashid and G. Winstone for discussions and P. Connell for expert technical assistance in setting up the experiments. A.S. is supported by the Engineering and Physical Sciences Research Council (EPSRC) under Centre for Doctoral Training Grant No. EP/L015382/1. We would also like to thank the Leverhulme Trust {[}RPG-2016-046{]} and the EU Horizon 2020 research and innovation programme under grant agreement No 766900 {[}TEQ{]} for funding support. All data supporting this study are openly available from the University of Southampton repository at https://doi.org/10.5258/SOTON/D1120.

\bibliographystyle{apsrev4-1}
\bibliography{references.bib}

\end{document}